\documentclass[12pt,thmsa]{article}
\usepackage{sw20lart}

\input tcilatex
\begin{document}

\author{Amjad Hussain Shah Gilani \\
National Center for Physics\\
Quaid-i-Azam University\\
Islamabad 45320, Pakistan\\
Email: ahgilani@yahoo.com}
\title{Mass relations among quarks and/or leptons}
\date{}
\maketitle

\begin{abstract}
The mass relations among respective quark family members are predicted and
similar mass relation for the lepton family member are obtained. The volume
of the volume element is also calculated which is $\sqrt{3/2}$.
\end{abstract}

\section{Introduction}

The theory of strong interactions is named: quantum chromodynamics (QCD) and
it is based on massless bi-colored gluons. The study of the properties of
the strong interactions in the asymptotic Bjorken limit has proved to be one
of the most creative ideas in theoretical physics \cite{hep-ph/0502113}.
Relatively little work has been done in the other high energy limit, i.e.
Regge limit \cite{hep-ph/0502190}. In Regge asymptotics, the number of
partons increases rapidly due to QCD bremsstrahlung. The massless eight
gluons carry color-anticolor charge \cite{Griffiths1987}. QCD, in
particular, is constructed as the unique embodiment of a symmetry group,
local SU(3) color gauge symmetry. To use symmetry as our guide, QCD plays an
essential tool in devising strategies for experimental exploration.
Unfortunately, a group property violation is observed in the color charge
structure of gluons \cite{hep-ph/0404026}. This destroys the whole structure
of the QCD. A new model for strong interactions is proposed which favored
electroweak and can be extended to the study of Casimir and gravitational
forces \cite
{hep-ph/0404026,hep-ph/0410207,hep-ph/0501103,hep-ph/0502055,hep-ph/0502117}%
. This model made predictions with the help of set theory and get
constraints from group theory. A unified picture of all the four forces
(i.e. Casimir force, gravitatonal force, electroweak force and strong force)
appeard and nature of the forces appeared to be electroweak.

Presently, we study the possible relation among the masses of quarks and/or
leptons families.

\section{The size of volume element}

Quantum Chromodynamics (QCD) has three charges i.e. red, green and blue. QCD
developed by giving color charges and fractional charges to quarks while the
color charges were arbitrary. No value was given to the color charges.
Recently, the value of color charges is predicted in Ref. \cite
{hep-ph/0410207} with the help of cube roots of unity whereas the fractional
charges of quarks are discarded. With the help of pictorial representation
of cube roots of unity, various mass relations among gluons are also
predicted. By keeping in view of the discussion of various figures of Ref. 
\cite{hep-ph/0410207}, we can exactly estimate the size of the volume
element. Let us have a look on Fig. 1, the ratio among the various line
elements are: 
\begin{eqnarray}
oz &:&or:\bar{g}\bar{b}=\sqrt{2}:1:\sqrt{3}  \nonumber \\
oz &:&or:\frac{\bar{g}\bar{b}}2=\sqrt{2}:1:\frac{\sqrt{3}}2  \label{e1}
\end{eqnarray}
As the point `$o$' is the center of the $\gamma $ (photon) volume element
and `$z$' is that of $Z^0$. So, the height of the volume element is $oz$.
Similarly, the length of the volume element is `$or$' and width is $\bar{g}%
\bar{b}/2$. Therefore, the volume of the volume element is 
\begin{equation}
V=\sqrt{2}\times 1\times \frac{\sqrt{3}}2=\sqrt{\frac 32}.  \label{e2}
\end{equation}

\section{Mass relations among quark families}

Quarks were known by three up-types (up, charm, top) and three down-types
(down, strange, beauty). Beside the color charges, quarks were supposed to
have electric charges whose magnitudes are fractions (2/3 or 1/3) of what
appears to be the basic unit, namely the magnitude of the charge carried by
proton or electron. Value to the color charges is given by cube roots of
unity \cite{hep-ph/0410207}. After giving the value to color charges, it is
pointed out that the fractional charges are useless and hence rejected.
Among six flavors of quarks, three are light (i.e. down, up, strange) and
three are heavy (i.e. charm, beauty, top). Instead of six flavor of quarks,
three quark families are proposed (i.e. charm, beauty, top) while the light
quarks (i.e. down, up, strange) are accomodated in these families \cite
{hep-ph/0501103}. The triplicity of quarks is discussed in Ref. \cite
{hep-ph/0502117} and suspected that three faces of a cube provide
information about quarks and three about leptons. The study of cube roots of
unity provide a ratio among the sides of the volume element [see Eq. (\ref
{e1})]. If we depict the same to the masses, as the charm is the lighest
among quarks and top the heaviest one. Therefore, 
\begin{equation}
m_{t_i}:m_{b_i}:m_{c_i}=\sqrt{2}:1:\frac{\sqrt{3}}2,
\end{equation}
where $i$ represent the family member of the respective family. Similar
relations can be obtained for leptons 
\begin{equation}
m_{\tau _i}:m_{\mu _i}:m_{e_i}=\sqrt{2}:1:\frac{\sqrt{3}}2.
\end{equation}

\section{Conclusions}

The mass relations among respective quark family members are predicted and
similar mass relation for the lepton family member are obtained. The volume
of the volume element is also calculated.

\textbf{Acknowledgements: } First of all, I thanks to all of my friends and
collegues for their comments and apology for late acknowledgement. Thanks to
Dr. D. V. Ahluwalia-Khalilova for his comments during April 2004 and a
question about the possibility of scalars in the theory. This help me to
figure out Higgs boson in the model \cite{hep-ph/0410207}. I requested
Professor Guido Altarelli for comments during April 2004, his criticism
helped me to think more deeply which results to write articles \cite
{hep-ph/0410207,hep-ph/0501103,hep-ph/0502055,hep-ph/0502117}. I am also
thankful to Dr. S. Basit Athar, Dr. Iqbal (UET), Dr. Khattak (Gomal
University), Shahzad (PAEC), Shafiq (Sweedon) for their comments and
criticism. I requested for comments to Dr. M. Mangano, his comments and
criticism are valueable and hope help me a lot to take step forward in the
right direction. Many thanks to Dr. Mangano in this regard. Many thanks to
Professor Frank Wilczek for a reply upon my request of comments. Finally I
thanks to Professor Guido Altarelli for discussions during the 3rd Workshop
on Particle Physics, 8--13 March 2004, Islamabad, Pakistan.

\section{Figure Captions}

\begin{enumerate}
\item  One of the possible plot of cube roots of unity \cite{hep-ph/0410207}.
\end{enumerate}


\begin{thebibliography}{9}
\bibitem{hep-ph/0502113}  F. Wilczek, Asymptotic freedom: from paradox to
paradigm, [hep-ph/0502113]

\bibitem{hep-ph/0502190}  R. Venugopalan, The color glass condensate: an
overview, [hep-ph/0502190]

\bibitem{Griffiths1987}  D. Griffiths, Introduction to elementary particles,
John Wiley \& Sons (1987)

\bibitem{hep-ph/0404026}  A. H. S. Gilani, Are gluons massive ?,
[hep-ph/0404026]

\bibitem{hep-ph/0410207}  A. H. S. Gilani, The value of color charges and
structure of gauge bosons, [hep-ph/0410207]

\bibitem{hep-ph/0501103}  A. H. S. Gilani, How many quarks and leptons ?,
[hep-ph/0501103]

\bibitem{hep-ph/0502055}  A. H. S. Gilani, Why does group theory fail to
describe charge structure of particles ? [hep-ph/0502055]

\bibitem{hep-ph/0502117}  A. H. S. Gilani, Why does set theory necessary to
describe charge structure of particles ? [hep-ph/0502117]
\end{thebibliography}
\end{document}